# Deep Learning Interatomic Potential Connects Molecular Structural Ordering to Macroscale Properties of Polyacrylonitrile (PAN) Polymer

*Rajni Chahal[1*], Michael D. Toomey[1], Logan T. Kearney[1], Ada Sedova[2], Joshua T. Damron[1], Amit K. Naskar[1], Santanu Roy[1*]*

[1]Chemical Science Division, Oak Ridge National Laboratory, Oak Ridge, TN-37830, United States

[2]Bioscience Division, Oak Ridge National Laboratory, Oak Ridge, TN-37830, United States

chahalr@ornl.gov, roys@ornl.gov

**ABSTRACT**

Polyacrylonitrile (PAN) is an important commercial polymer, bearing atactic stereochemistry resulting from nonselective radical polymerization. As such, an accurate, fundamental understanding of governing interactions among PAN molecular units are indispensable to advance the design principles of final products at reduced processability costs. While *ab initio* molecular dynamics (AIMD) simulations can provide the necessary accuracy for treating key interactions in polar polymers such as dipole-dipole interactions and hydrogen bonding, and analyzing their influence on molecular orientation, their implementation is limited to small molecules only. Herein, we show that the neural network interatomic potentials (NNIP) that are trained on the small-scale AIMD data (acquired for oligomers) can be efficiently employed to examine the structures/properties at large scales (polymers). NNIP provides critical insight into intra- and interchain hydrogen bonding and dipolar correlations, and accurately predicts the amorphous bulk PAN structure validated by modeling the experimental X-ray structure factor. Furthermore, the NNIP-predicted PAN properties such as density and elastic modulus are in good agreement with their experimental values. Overall, the trend in the elastic modulus is found to correlate strongly with the PAN structural orientations encoded in Hermans orientation factor. This study enables the ability to predict the structure-property relations for PAN and analogs with sustainable ab initio accuracy across scales.

**INTRODUCTION**

Polyacrylonitrile (PAN) is an important industrial polymer used in a wide variety of commercial products including membranes [1], [2], textiles [3], [4] and carbon fiber [5] due to its high thermal stability and good mechanical properties. In PAN, due to the presence of polar C≡N pendant groups, strong dipole-dipole interactions exist [6], which influence the local chain configurations as well as the bulk macroscopic polymer structure. Additionally, the selection of polymerization process can introduce deviation in molecular weight and stereoregularities in PAN chains [7], [8], which, in turn, can influence their structure and resulting properties [9], [10], [11], [12]. As such, a fundamental understanding of the effect of governing interactions on chain configurations for such systems can assist with tuning processability parameters, which could eventually lead to lower production costs while achieving better final product quality. Furthermore, depending on its conformation and orientation factor, PAN macromolecules exhibit potential to chemically transform and yield flame resistant products with excellent mechanical properties along with chemical and oil resistance [5]. An enhanced understanding of these factors on the structure and properties of PAN can be further useful towards understanding of other similar commercial polymers, such as polyvinylchloride, which exhibits similar dipole-dipole interactions due to the presence of the electronegative chloride group.

In the past, several experimental studies garnered an understanding of PAN bulk structure using X-ray diffraction [7], [13], [14], [15] and nuclear magnetic resonance (NMR) [8], [16]. The PAN chain conformations, as well, have been long studied in detail using the perturbed configuration interaction with localized orbitals (PCILO) method [7], [17]. Additionally, atomistic modeling and molecular dynamics (MD) using classical force fields, such as OPLS [18], DRIEDING [19], etc. have been successfully used to predict their bulk structure [10], [18], [20], [21], glass transition temperature [22] and densities [10], [18]. Despite commendable agreement in the structure prediction between X-ray diffraction data [10] and the MD efforts made by Shen et al. [10], the elastic properties evaluated in their study remain overpredicted by more than 100% when compared to experimental values. This is likely due to a less-accurate representation of electrostatic interactions, especially between C≡N pairs. As such, a large contribution of electrostatic interactions between C≡N pairs represented in the available force-fields lead to more-attractive interactions in PAN, which eventually lead to the observed deviation in the calculated properties [10].

The density functional theory (DFT)-based *ab initio* molecular dynamics (AIMD) simulations are well-suited for treating polarization effects, charge transfer, and hydrogen bonding in condensed-phase systems [23]. However, their higher computational cost imposes limitations on the system size (< 20 monomer units) as well as simulation times (picoseconds). Therefore, attaining ergodicity and statistically converged macroscale properties with AIMD is infeasible. To this end, the emerging machine learning (ML) potentials-based MD simulations may prove promising as these ML potentials can be directly trained on the AIMD data and can then be used to treat molecular interactions across scales with *ab initio* accuracy [24], [25], [26], [27]. Recently, Gaussian approximation ML potential (GAP) has been used to study the structure and dynamics of simple hydrocarbon systems, where their extensibility towards simulating long hydrocarbon chains was demonstrated [28], [29]. Although no properties were reported in this study, motivated by this effort, Mohanty et al. [30] developed a charge recursive neural network (QRNN) ML potential to study the structure, transport, and thermophysical properties of a simple polymer system. Here as well, the mechanical properties evaluation was not attempted. It should be noted that such charge-based potentials demand extensive computation and significant training efforts for complex systems such as PAN due to a multitude of calculations of dipole moments and atomic charges. As such, a highly parametrized neural network interatomic potential (NNIP) [25] can accurately map the potential energy surface of soft condensed-matter systems including PAN without requiring significant user intervention. Previously, such NNIPs have been shown to accurately predict the structure and properties of many complex polarizable liquids [31], [32], [33], [34], [35].

In this work, we developed and deployed an NNIP potential for PAN and provided a computational framework for extending the capability of NNIP development for polymers. Quite interestingly, while the NNIP was trained on the small-scale AIMD data generated for oligomers comprising up to 20 monomer units and their ensembles for picoseconds, the deployment of the NNIP through the neural network molecular dynamics (NNMD) was successful and efficient at large scales (> 300 monomer units and tens of nanoseconds), allowing us to thoroughly examine the density, structure, relaxation dynamics, and mechanical properties of PAN. Most-importantly, critical insight into the chain length-dependent probable PAN configurations (e.g., folded or partially-folded structures stabilized by H-bonds/dipolar interactions and unfolded structures) and PAN chain orientational ordering, and their correlation with the macroscale properties such as elasticity was gained. For validation, in addition to utilizing already-available experimental data [7], [36] (namely, X-ray scattering, dipole relaxation time, densities), we performed new experiments to determine elastic properties using the wet-spun PAN fibers prepared at Oak Ridge National Laboratory. The findings of NNMD agree well with these measured properties. To compare with the performance of a reference force field, we chose ReaxFF [37] which is recently and extensively used for several PAN-type systems [38], [39], [40], [41], [42], [43], and found that in addition to maintaining *ab initio* accuracy, NNIP is advantageous over ReaxFF in terms of computational efficiency. The satisfactory experimental validation and attained computational efficiency and scalability provide confidence in the extendibility and further deployment of NNIP for large polymeric systems.

## 2. METHODS

In this section, we discuss the computational and experimental methods used to determine the PAN structure, dynamics, and properties. First, we present AIMD data generation and their utilization in the NNIP development and provide the details of the NNMD and ReaxFF simulations. Then, we define key collective variables for structural and dynamical analyses. Finally, we discuss the materials and their characterization in experiments conducted.

## 2. 1 Computational Methods and Structure Characterization

### Neural Network Interatomic Potential (NNIP) Development

*Data generation: Ab-Initio Molecular Dynamics Simulations*

To generate NNIP training data, Born-Oppenheimer AIMD simulations were performed using the QUICKSTEP module of the CP2K package [23], the double-zeta MOLOPT basis sets (DZVP-MOLOPT) with a density cutoff of 400 Ry, and Goedecker-Teter-Hutter (GTH) pseudo-potential. The Perdew, Becke, and Ernzerhof (PBE) functional was used to treat the exchange-correlation interactions. Grimme's D3 correction to the dispersion interactions was applied [23], [44]. The canonical ensemble (NVT) using a Nosé-Hoover chains thermostat [45] with 100 fs time constant was employed while maintaining the periodic boundary conditions. Several AIMD simulations were performed using 1 fs timestep for 1.5-27 ps, specific details for which are provided in the next section.

*NNIP Training*

For training of interatomic potential, DeePMD-kit (DP-kit) package (version 2.1.1) was employed [46]. The Deep-Pot-Smooth Edition (DeepPot-SE) potential contained inside the DP-kit was chosen due to smooth and continuously differentiable potential energy surface generation [25]. To train NNIP for PAN, the training dataset comprised 27,324 10 monomer unit (78 atoms) and 4,368 20 monomer units (148 atoms) configurations at 1000 K, which were end-capped with $CH_3$ group. Here, this extremely high-temperature was chosen for a very short AIMD simulation time (< 30 ps), which allowed rapid and extensive sampling of the PAN configurations but did not degrade the oligomers. However, due to strong dipole-dipole interactions existing in PAN, including the *ab initio* data acquired for the isolated PAN configurations sampled at the high temperature in the training alone were not sufficient to run stable NNMD for a longer time scale during which the longer chains may fold and interact with its segments. As such, we found that in these cases, including relatively fewer bulk PAN configurations significantly improved the robustness and reliability of the NNIP when employed to simulate longer isolated PAN chains. In specific, 1722 bulk (comprised of four 10-mer PAN) configurations at 300 K at experimental density were included in the training. Further, to ensure that the NNIP remains stable and has the density prediction capability, the bulk configurations from same PAN system with varied densities were also included in the training. Specifically, 1580 bulk configurations at 5% compressed, 2013 at 11% expanded, and 2149 at 34% expanded volumes (relative to experimental densities) were included in the NNIP training. It is worth noting that the need for including compressed and expanded configurations in addition to the relaxed (near zero pressure) AIMD configurations to obtain stable densities from NNIP was also pointed out in our and other previous studies [31], [33]. During the training, the datasets were shuffled and split, with 80% and 20% representing the training and validation sets, respectively. Overall, the training data was selected based on training-validation-augmentation procedures previously used to create robust NNIPs that can accurately predict structures and common thermophysical properties of other complex systems [31].

During training, the DeepPot-SE model (see Supporting Information (SI) for details) learns a mapping between the local environment of each atom within 8 Å cut-off and a per-atom energy, such that the sum of atomic energies corresponds to reference DFT energy. The gradients of the NNIP-predicted energies are then used to compute the atomic forces. Both the reference energies and forces are included to evaluate the loss function which is minimized during training of an DeepPot-SE model. Here, the smooth cutoff and hard cutoff radius of 2 Å and 8 Å is chosen. The embedding network and fitting network sizes are {25,50,100} and {400,400,400}, respectively. The tunable prefactors in the loss function were chosen as 0.002, 1000, 1, 1 for $p_e^{start}$, $p_f^{start}$, $p_e^{limit}$, and $p_f^{limit}$, respectively. These hyperparameters previously showed success in predicting accurate structure and transport properties of a multicomponent system in our previous study [31]. Herein, the trained network using DeepPot-SE yielded average energy and force errors of 0.925 meV/atom and 76.2 meV/Å, respectively for the validation set. These errors are within the precision of DFT energies/forces. As such, the low energy and force testing errors suggest a well-fitted potential energy surface.

**Neural Network Molecular Dynamics**

The trained NNIP potential was used in LAMMPS [47] *via* the interface with DeePMD-kit [47]. Here, the transferability of NNIP towards longer polymeric chains was tested by studying both isolated chains and bulk PAN systems where each PAN chain contained 10-320 monomer units. A bulk PAN system comprising infinitely periodic PAN chains (no-end $CH_3$ groups, i.e., a continuous PAN chain) was also considered for NNMD simulations. To create the latter system, 160-mer chains were aligned along x-direction while maintaining periodicity of chain ends (**Figure S1**). The remaining bulk systems were prepared by randomly placing n-mer (n: 10-320) PAN chains in large box sizes using Packmol [48]. A detailed description of initial box sizes and corresponding snapshots are provided in **Figure S2**. The equilibrated simulation cell details for both chains and bulk PAN systems are reported in **Table 1**. During NNMD, isolated PAN chains systems were relaxed for at least 5 nanosecond (ns) at 300 K using Nosé-Hoover thermostat [49] with a 100 fs time constant. After this step, a 6-10 ns trajectory was generated for each n-mer system to study the radius of gyration ($R_g$), end-to-end root mean squared distance ($RMSD_{end-to-}$ $_{end}$), and H-bonding in equilibrated PAN chains. **Figure 1** and **Figure 2** show the initial and relaxed PAN configurations, respectively.

**Table 1.** Details of PAN systems studied using AIMD, ReaxFF, and NNIP simulations.

| n-mer units | Chains (ReaxFF, NNIP) | Bulk (NNIP) |
| --- | --- | --- |
| 10 | 78 atoms, 60 × 60 × 60 Å | 16,848 atoms, ~92 × 46 × 46 Å |
| 20 | 148 atoms, 60 × 60 × 60 Å | 29,600 atoms, ~127.5 × 51 × 51 Å |
| 30 | 218 atoms, 60 × 60 × 60 Å | 43,600 atoms, ~145.9 × 58.4 × 58.4 Å |
| 40 | 288 atoms, 60 × 60 × 60 Å | 34,560 atoms, ~96 × 64 × 64 Å |
| 80 (suffix: ex1) | 568 atoms, 100 × 100 × 100 Å | 22,720 atoms, ~95.5 × 63.7 × 63.7 Å |
| 80 (suffix: ex2) | 568 atoms, 200 × 200 × 200 Å | 22,720 atoms, ~101.7 × 50.8 × 50.8 Å |
| 80 (suffix: ex3) | 568 atoms, 200 × 200 × 200 Å | - |
| 160 (suffix: ex1) | 1128 atoms, 120 × 120 × 120 Å | 67,680 atoms, ~120.4 × 80.2 × 80.2 Å |
| 160 (suffix: ex2) | 1128 atoms, 300 × 300 × 300 Å | - |
| 320 | 2248 atoms, 520 × 520 × 520 Å | 31,472 atoms, ~115.4 × 55.7 × 55.7 Å |
| Continuous chain | - | 28,000 atoms, ~217 × 39 × 39 Å |

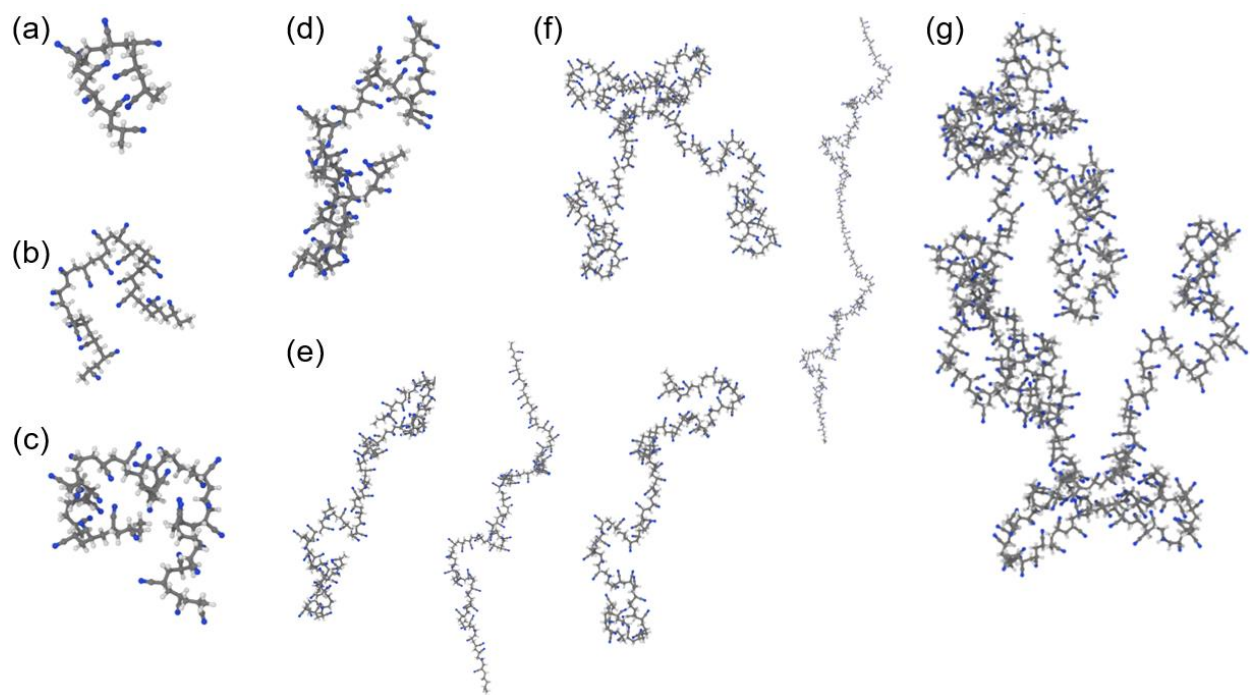

**Figure 1.** Before equilibration PAN configurations generated using open-access geometry module in Materials Square [50] containing (a) 10, (b) 20, (c) 30, (d) 40, (e) 80, (f) 160, and (g) 320 monomer units. Three initial configurations were considered for 80-mer and two initial configurations were considered for 160-mer PAN chains.

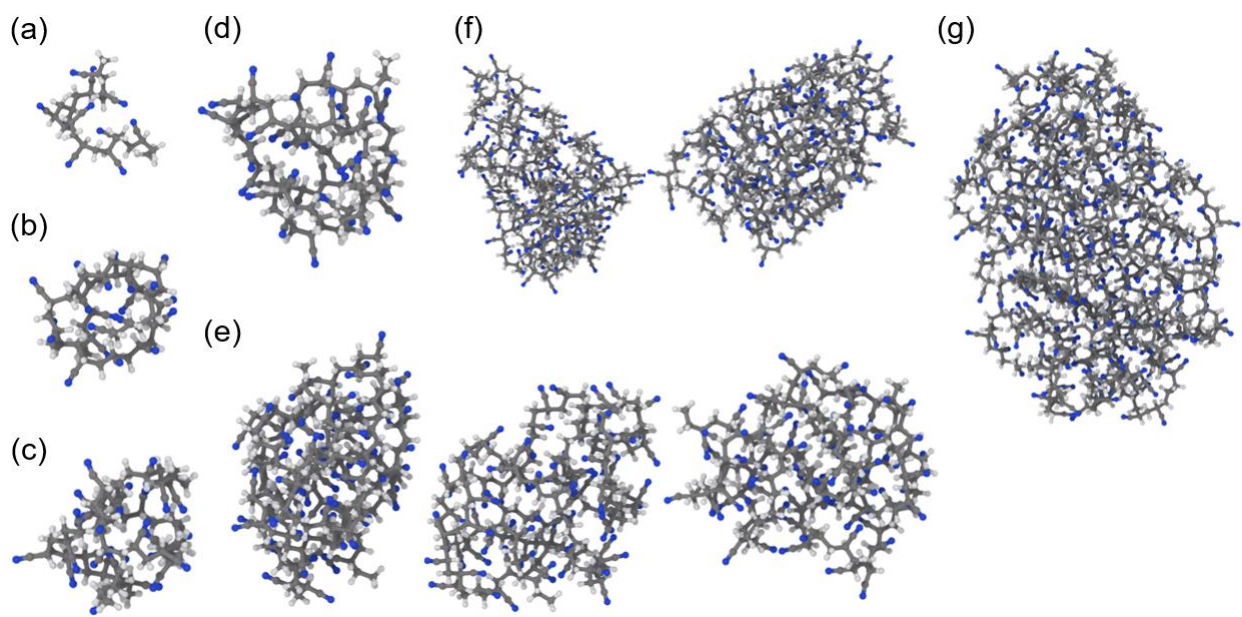

**Figure 2.** Relaxed PAN configurations as obtained after NNIP-based MD simulations for (a) 10, (b) 20, (c) 30, (d) 40, (e) 80 (ex1, ex2, ex3), (f) 160 (ex1, ex2), and (g) 320 monomer units.

For bulk PAN systems (shown in **Figure S2**), the systems were compressed for nearly 2-4 ns at 300 K using NVT ensemble to achieve average box densities ~1 g/cc. Thereafter, an isothermal-isobaric ensemble (NPT) along with the Nosé-Hoover thermostat and barostat [49] was used to relax the volumes to achieve respective equilibrium densities at atmospheric pressure and 300 K conditions. Afterwards, ~5 ns simulations at equilibrium densities were performed using NVT along with Nosé-Hoover thermostat [49] to relax the system at corresponding densities. The equilibrated trajectory of more than 3 ns was used to evaluate the density, radial distribution function, and hydrogen bonding (H-bonding) in bulk PAN for all 10-, 20-, 30-, 40-, 80-, 160-, 320-mer, and continuous PAN chain systems. Further, the NNMD simulations were extended for ~10-15 ns for the 10-mer, 160-mer, and continuous bulk PAN systems to predict dipole relaxation behavior in bulk PAN. All bulk systems were considered for evaluating mechanical properties using the developed NNIP potential. To ensure proper load transfer among chains during deformation, the lateral simulation cell parameter (reported in **Table 1.**) was kept well above the previously considered cell lengths of ~30 Å in the literature [10]. For all PAN chains and bulk PAN simulations, a 1 fs NNMD timestep was chosen while maintaining the periodic boundary conditions. All NNMD simulations were performed at 300 K.

**Molecular Dynamics using ReaxFF Force Field**

CHON-2019 ReaxFF parameters developed by Kowalik et al. [37] were used to study isolated PAN chains with 10-320 monomer units. LAMMPS is employed for molecular simulations [47]. Using a 1 fs time step, at least 15 ns simulations were performed employing canonical ensemble (NVT) using Nosé-Hoover thermostat [49] with a 100 femtosecond (fs) time constant while maintaining the periodic boundary conditions. The simulation system details for PAN chains studied using ReaxFF are reported in **Table 1**. For all systems, nearly 6-10 ns of equilibrated trajectory was used for sampling PAN chain configurations to evaluate the radius of gyration ($R_g$), end-to-end root mean squared distance ($RMSD_{end-to-end}$), Flory's random coil relation, and H-bonding in equilibrated PAN chains at 300 K. These quantities are compared to NNIP's predictions.

**Characterization and Analysis of Simulated Structure**

*Coordination Number: An approximation for hydrogen bond numbers*

To calculate atomic pairs (e.g., C≡N….H) coordination in simulations, the smoothly varying coordination number [31], [51], [52] as given by **Equation 1** is used, where $r_i$ is the distance between the $i^{th}$ H and N of the C≡N group, $N_H$ is the total number of H atoms, and $r_{cut}$ represents the first minimum in the corresponding atomic pair distribution function, *g(r)*.

$$\boldsymbol{CN = \sum_{i=1}^{N_H} \frac{1-\left(\frac{r_i}{r_{cut}}\right)^{12}}{1-\left(\frac{r_i}{r_{cut}}\right)^{24}}} \quad (1)$$

Typically, a hydrogen bond is defined using geometric criteria based on distance and angle cutoffs between the hydrogen donor and acceptor molecules. Here, CN represents an approximate hydrogen bond number considering the distance cutoff obtained from calculated g(r).

*Dipole Relaxation Behavior*

To evaluate the dipole relaxation constant of PAN, a block-averaging scheme was employed. In this approach, the equilibrated simulation trajectory is divided into multiple blocks of length equal to correlation time (here, 5 ns). At t=0, each C≡N pair orientation ($\vec{U}(t_{0i})$ vector) is evaluated, which is substituted to **Equation 2** along with its updated orientation vector $\vec{U}(t+t_{0i})$ as time reaches correlation time.

$$ACF = \frac{1}{n_{C\equiv N}} \frac{1}{n_t} \sum_{j=1}^{n_{C\equiv N}} \sum_{i=1}^{n_t} \frac{1}{2}\left(3[\vec{U}(t+t_{0i}).\vec{U}(t_{0i})]^2 - 1\right) \qquad (2)$$

Here, multiple blocks ($n_t$) are considered by setting reference frames (t=0) throughout the equilibrated trajectory in order to obtain better averaging. This calculation is performed for all C≡N pairs in the system. In brief, the dipole autocorrelation function (ACF) was evaluated for each C≡N pair in the system using the block-averaging method. Finally, to obtain the average dipole ACF function, the calculated ACF values for each C≡N pair were averaged (**Equation 2**). When ACF value approaches zero, the C≡N bond loses the memory of its initial orientation and so the C≡N dipole decorrelates or relaxes fully.

*Simulated X-ray structure Factor (S(Q))*

The simulated X-ray S(Q) is obtained from the X-ray raw scattering intensities utilizing the Debye formula [53], [54] and employing the Debyer [53] and PDFgetX2 [55] softwares. The details are given in the SI (specifically, **Equations S3 and S4**).

*Hermans Orientation Factor (HOF)*

To quantitatively evaluate the backbone orientation of predicted PAN structure in NNIP MD simulations, Hermans orientation factor (HOF) was established as a parameter [56]. **Equation 3** was used to evaluate the HOF of C-C backbone chains, where $\theta_x$, $\theta_y$, and $\theta_z$ are the angle between C-C bonds in backbone chain with respect to the global x-, y-, or z-direction, respectively. As x-axis was chosen as the loading axis, only $\theta_x$ were considered in **Equation 3** to evaluate HOF with respect to loading direction. As such, an HOF value of 1 would represent a perfect orientation of backbone chain with respect to the loading axis, while 0 HOF suggests net random orientation of backbone C-C bonds. Conversely, -0.5 value for HOF suggests 90° orientation (perpendicularly aligned) relative to loading axis.

$$\text{HOF} = \frac{1}{2}(3\langle cos^2\theta_x \rangle - 1) = \begin{cases} -0.5 & (perpendicular) \\ 0 & (random) \\ 1 & (aligned) \end{cases} \quad (3)$$

**Scaling and Efficiency of NNIP compared to ReaxFF Simulations**

To examine and compare the computational demand of the NNIP with referenced ReaxFF potential, scaling tests were performed for bulk PAN, and the corresponding simulation times were noted. For this purpose, parallel installation of LAMMPS **[47]** utilizing 6 NVIDIA V100 GPUs per compute node for both NNIP and ReaxFF MD simulations was utilized on Summit cluster **[57]** in the Oak Ridge Leadership Computing Facility (OLCF).

**Figure 3** shows the computational demand of neural networks as compared with ReaxFF simulations, both containing a 28,000 atom bulk PAN system. The required core-hours to run 10,000 time steps of molecular dynamics simulation were evaluated. For both NNIP and ReaxFF, the computing time exponentially reduces to run 10,000 MD steps indicating good scaling of the system with increasing compute nodes. For all cases, the NNIP performs the same calculation in nearly half the CPU-hours, relative to the ReaxFF. Overall, the developed NNIP is found to be twice as efficient as ReaxFF, while maintaining *ab initio* accuracy. Therefore, in this study, while we compare between the results of specific structural analysis obtained from ReaxFF and NNIP, the only-NNIP simulations were considered for the expensive calculations such as the determination of thermophysical and mechanical properties of the bulk PAN.

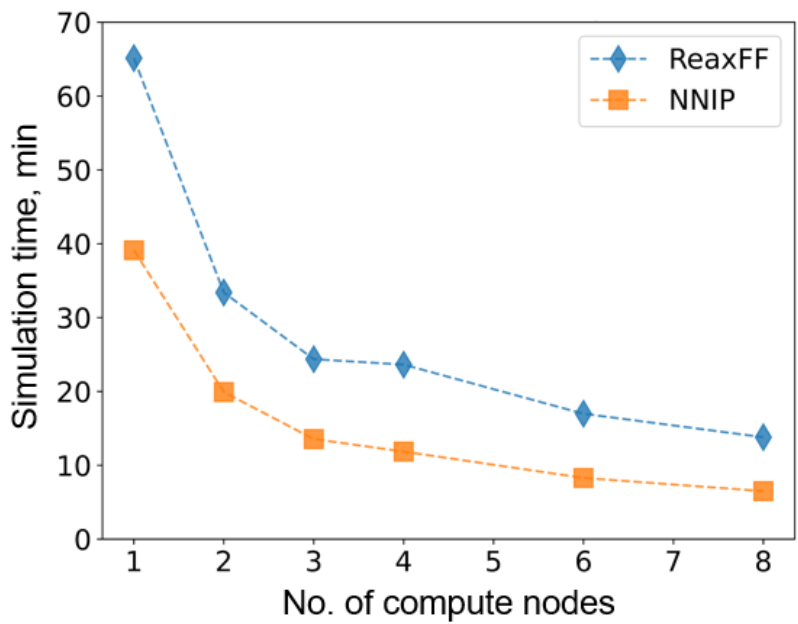

**Figure 3.** NNIP's performance and scalability compared to ReaxFF. Simulation time was evaluated to run 10,000 MD steps on 28,000 atoms system on Summit Supercomputer [57] (6 NVIDIA V100 GPUs per compute node).

**2.2 Experimental Materials and Methods**

Poly(acrylonitrile) (PAN; $M_w$=150,000 (for comparison, $M_w$=17,010 for 320-mer simulated model)) and dimethyl sulfoxide (DMSO; > 99.8%) were acquired from Sigma Aldrich and were used as-received. The considered molecular weight representing more than 2800 PAN monomer units may be comparable to the bulk continuous chain model among simulated systems. In experiments, PAN fibers were made through fiber spinning of a 12 wt.% solution of PAN in DMSO to achieve oriented samples of various molecular alignments. Oriented PAN fibers were spun using a 500-hole spinneret and a custom wet-spinning line comprised of four baths. By volume, bath compositions had 80-20, and 50-50 DMSO-water ratios for the first two baths, followed by one room temperature and one 98°C type-II water bath for the final two. During spinning, a jet stretch of 2 and a hot stretch of 4 were performed in the first and last baths to achieve an overall draw ratio of 8. After exiting the final bath, PAN fibers were coated with a spin-finish solution and annealed at 120°C and 150°C on successive hot annealing rollers before being collected on a take-up winder. Three separate samples were collected using this method. After initial spinning, a steam stretcher was used on two of the three samples to attain steam-relaxed and steam-stretched samples. For the steam-relaxed sample, the PAN tow was passed through the steam-stretching unit and exposed to steam for 120 seconds to enable polymer chain relaxation within the tow. This was performed statically, using a 0.05 kg weight to prevent the tow from sagging, while still allowing for some retraction. For steam-stretched samples, the process was repeated, using a 5 kg weight to stretch the fiber tow, further increasing polymer chain alignment. The three samples, referred to as steam relaxed (SR), as-spun (AS), and steam stretched (SS) were then characterized for internal structure and mechanical performance.

**Structural characterization**

*Experimental validation of key physical properties*

Cross-validation of the NNIP-predicted PAN structure was performed using highly oriented PAN fibers generated in our lab. Furthermore, we combined mechanical property measurements along with structural characterization via x-ray scattering to establish structure-property correlations in our materials. Notably, the NNIP provides a molecular picture of a densified PAN bulk without the presence of crystallites. As a result, HOF extracted from the simulated cell results from the distribution of vectors which constitute the PAN chain backbone (**Equation 3**). Experimentally, x-ray scattering is sensitive to differences in electron density and therefore increases in density within the crystal volumes lead to strong scattering centers. This difference in how HOF is derived computationally and experimentally precludes direct comparison between the values, however, qualitative relationships remain valid between the two approaches. Atactic PAN exhibits a disordered helical chain structure and the resulting crystals form pseudohexagonal crystallites. The primary (100) crystal plane is aligned along the fiber axis and results in scattering features indicative of the variation in orientation. The scattering intensity vs. azimuthal angle resulting from the (100) plane in the PAN fibers is collected via wide-angle x-ray scattering. Since the PAN backbones within the crystallites are highly registered and aligned with the backbone relative to the surrounding amorphous regions, we expect a much higher HOF.

Microstructural characterization of the PAN fibers and films were performed with a Xeuss 3.0 (Xenocs, France) equipped with a D2+ MetalJet X-ray source (Ga Ka, 9.2 keV, $\lambda$ = 1.34 Å). 2D images of the scattering patterns were collected on a PILATUS hybrid photon counting detector with a pixel dimension of 75 x 75 $\mu m^2$ (Dectris, Switzerland). The 2D SAXS images were circularly averaged and reduced in the form of absolute intensity versus scattering vector (q), where q = ($4\pi \sin \theta$)/$\lambda$. All images were collected with background corrections and reduced using the XSACT software package (Xenocs, France). 1D azimuthal profiles were acquired from the reflections of the (100) plane. Reduced intensity vs scattering vector (q) were plotted based on the 0 and 90 ° (± 2.5°) directions. Instrumental broadening was removed for calculations based on the effects from Scherrer broadening by progressive slit reduction on a crystalline standard. Orientation of the (100) in the axial direction was quantified with Hermans orientation factor (*f*) was calculated according to **Equation 4** and **5** where $\emptyset$ is azimuthal angle, and *I*($\emptyset$) is the scattering intensity:

$$f = \frac{3<\cos^2\emptyset>-1}{2} \quad (4)$$

$$<\cos^2\emptyset> = \frac{\int_0^{\pi/2} I(\emptyset)\cos^2\emptyset \sin\emptyset \, d\emptyset}{\int_0^{\pi/2} I(\emptyset) \sin\emptyset \, d\emptyset} \quad (5)$$

**RESULTS & DISCUSSION**

**Validating NNIP for PAN Chains**

*Radius of Gyration, RMSD*

The effect of dipole-dipole interactions in PAN on the chain relaxed structure can be estimated by studying the radius of gyration ($R_g$), which typically is a useful quantity when comparing the different solvents for PAN processing. Here, the effect of chain n-mer units on $R_g$ was studied using more than 6 ns trajectory obtained from both ReaxFF and NNIP potential MD simulations (**Figure 4(a)**). It is found that both ReaxFF and NNIP predicted similar values as well as the trend in $R_g$ as n-mers increase. However, a discrepancy in the $R_g$ values is observed as n-mer count increases to > 300 units. This can be attributed to a larger deviation in $R_g$ values as the initial chain configurations vary. For instance, it is observed that considering different initial chain configurations (shown for 80 n-mers, 160 n-mers) can influence the $R_g$ values of equilibrated chain structure due to different local chain folding as n-mer count increases. This is because of strong dipole interactions in PAN chains which can greatly influence the local chain morphology, especially when the molecular weight (i.e., n-mer units) of the PAN chain increases. Further, the end-to-end chain distance measured between $CH_3$ end groups ($RMSD_{end-to-end}$) was also evaluated for both ReaxFF and NNIP calculations. As shown in **Figure 4(b)**, the relation between the sum of square of $R_g$ is established with the sum of the square of $RMSD_{end-to-end}$ for all n-mer units using

ReaxFF as well as NNIP potential. The dashed red line shows Flory's random coil hypothesis/theoretical relation, which states $<\text{RMSD}^2_{\text{end-to-end}}> = 6<R_g^2>$ [58]. Typically, Flory's hypothesis holds well for freely jointed, non-interacting polymer chains [58]. As for smaller n-mer units, dipole-dipole interactions are comparatively less dominating, the NNIP predicted relation agreed quite well with the free jointed chain model. A similar trend is observed in ReaxFF predictions, which suggests a much smaller proportionality constant than that predicted from NNIP and theoretical relation. Here, a much larger value is predicted for the 320 n-mer unit system, which could also be due to limited sampling, as both $R_g$ and $\text{RMSD}_{\text{end-to-end}}$ values for large n-mer chains can have higher uncertainty depending on starting initial configurations.

Overall, for smaller chains (< 80-mer units), the $R_g$ and $\text{RMSD}_{\text{end-to-end}}$ relation predicted by NNIP agrees well compared to the theoretical relationship. However, their relation deviates from linear trend as chain n-mer units increase, indicating smaller values of $\text{RMSD}^2_{\text{end-to-end}}$ for a given $R_g^2$ value. This can be attributed to the strong dipole-dipole interactions in PAN as the n-mer count increases. In specific, as the n-mer units in the case of highly interacting chain (e.g., PAN) increase, the $R_g$ becomes smaller due to large attractive forces (between C≡N dipoles in PAN), and hence the proportionality constant is expected to decrease (<6). A similar deviation from Flory's hypothesis was also observed for PAN chain comprising 150 n-mer units as studied using OPLS parameters [18]. Typically, for such cases where significant interactions exist between the chains, a modification from Flory's hypothesis is proposed in [59], [60], [61]. **Figure 5** shows the relation between $R_g$ and the number of backbone C atoms ($N$), which is in the form of $R_g \sim N^x$. For instance, for globular protein systems, where strong chain interactions exist, *x~0.380*.

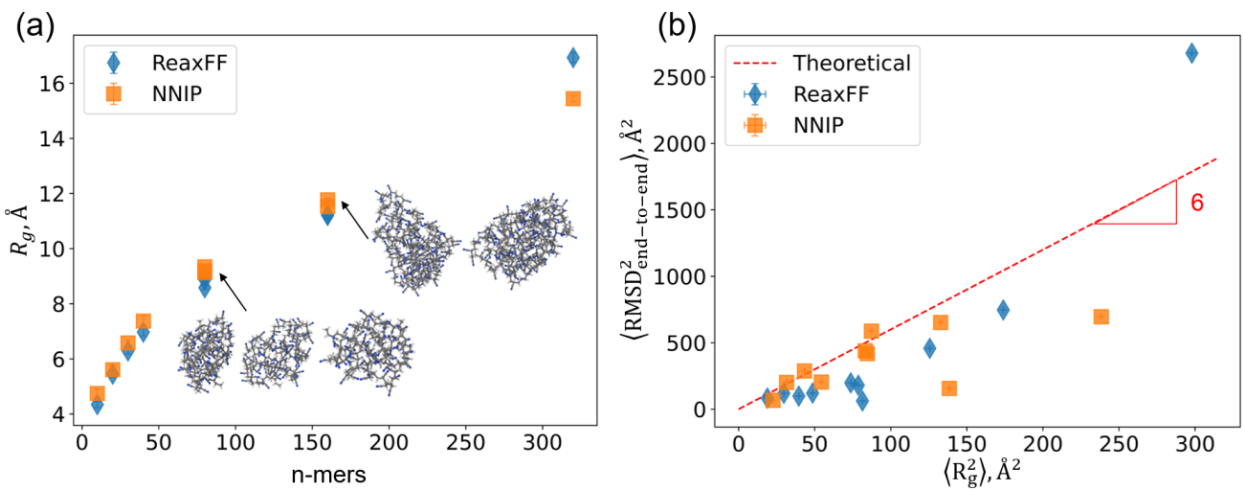

**Figure 4.** ReaxFF and NNIP predictions of (a) radius of gyration/$R_g$ and (b) relation between $R_g$ and RMSD$_{\text{end-to-end}}$ in comparison with Flory's theoretical relation for PAN chains as a function of n-mers. Relaxed chain configurations are shown in the figure inset.

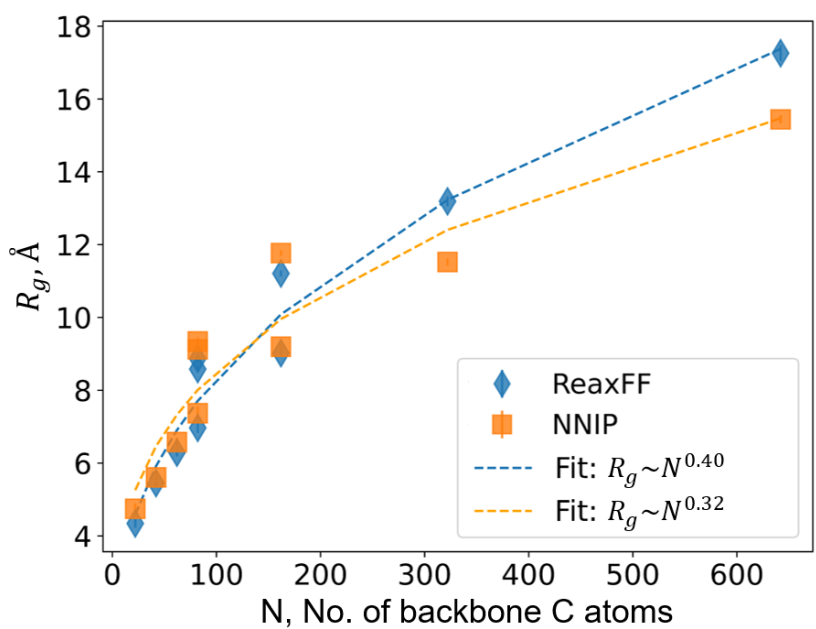

**Figure 5.** ReaxFF and NNIP predictions of relation between $R_g$ and No. of backbone atoms ($N$) for PAN chains as a function of n-mers. $R_g \sim N^\alpha$ relation exists for highly interacting chains [59], [60].

*H-bonding*

To further quantify the PAN chain structure as simulated using ReaxFF and NNIP potential, H-bonding interaction is approximated using C≡N….H coordination for each n-mer chain system (**Figure 6**). In order to calculate C≡N….H coordination, the smooth coordination number as given by **Equation 1** (see **Methods**) is used, where $r_{cut}$ value was chosen based on the first minimum in H-N partial g(r) observed at ~2.85 Å (Figure 6(a)). **Equation 1** allows smooth transitions of H across the boundary of the first C≡N coordination sphere, providing the instantaneous value of CN throughout the MD simulation. For comparison, the calculated value of CN for each system was normalized by the number of n-mer units in a chain (here, also equal to the number of C≡N pendant groups in a chain). As shown in **Figure 6(b)**, ReaxFF consistently predicts higher numbers of H-bonds/n-mer for >10 n-mer systems when compared to the NNIP-predicted values. This is due to excessive C≡N….H coordination as observed from the first peak at ~1.65 Å in ReaxFF simulations. This first peak is consistently observed in the ReaxFF simulations, even for 10-mer chain systems, while it is absent in both NNIP and AIMD simulations for the same system (**Figure S3**). This suggests the presence of stronger electrostatic interactions in the ReaxFF simulations (also observed in [10]), which eventually lead to higher values of H-bonding/n-mer as observed in **Figure 6(b)**. Nevertheless, here, both ReaxFF and NNIP suggest an increase in the net H-bonding/n-mer as the PAN n-mer units increase. Again, this can be explained from the perspective of strong dipole-dipole interactions in PAN, which leads to a deviation in $R_g$ trend as the chain molecular weight (i.e., n-mer count) increases (**Figure 4(a)**). As such, large C≡N pairs (n-mer units) per chain lead to higher attractive forces and tighter packing of chain segments, which correspond to an increased net contribution of H-bonds as observed in **Figure 6(b)**. Both ReaxFF and NNIP simulations predict that a plateau is achieved for n-mers > 80, indicating a saturation in hydrogen bonding.

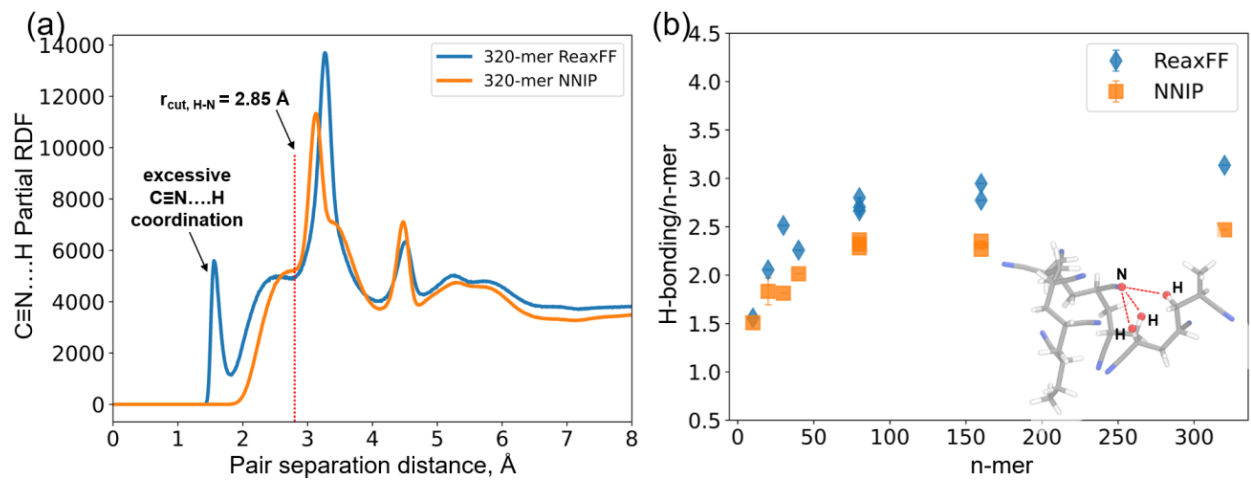

**Figure 6.** (a) C≡N….H Partial RDF for 320-mer chain and (b) H-bonding approximated using C≡N…..H coordination as a function of n-mer for PAN chains in the ReaxFF-based and NNIP-based MD simulations. ReaxFF shows higher C≡N….H coordination due to more contribution from electrostatic interactions. The inset shows a snapshot of intrachain C≡N…..H coordination of 3 in the 10-mer PAN chain.

**Deploying NNIP Potential for Bulk PAN**

*Density calculation*

The densities obtained for the bulk PAN systems containing continuous PAN chains and 10-320 monomer units PAN as evaluated from ~3 ns NNMD simulations are reported in **Figure 7**. The bulk density is sensitive to the n-mer units when n-mer count is < 40 units. As the n-mer count increases, the bulk densities achieve a constant value of ~1.1±0.1 g/cc. Here, the NNIP is able to predict the bulk PAN densities within 7% of the reported experimental values of 1.17-1.22 g/cc [62]. A slight underprediction of densities compared to experimental values may be due to inherent errors in training data caused by the choice of the DFT functionals and empirical dispersion corrections (PBE-D3). Additionally, the exclusion of long-range effects in NNIP training outside of the cut-off radius of 8 Å may also contribute to the small deviation in density values [32]. Also, typically in experiments, the PAN structure obtained comprises both amorphous and crystalline phases [21], whereas due to limited simulation cell sizes in MD simulations, the representation of crystalline phases is challenging [10]. As such, the presence of crystalline phases in experimentally measured PAN increases the average PAN density [20], which obviously can be higher than what the NNMD simulations predicted.

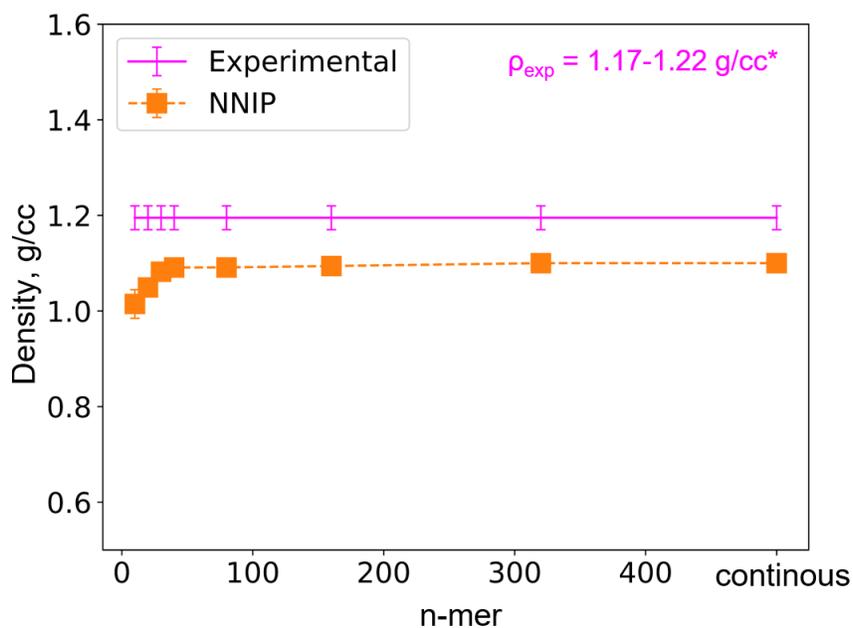

**Figure 7.** PAN density prediction by NNIP.

***Structural Analysis using Radial Distribution Function (RDF), X-Ray Structure Factor (S(Q))***

The developed NNIP was further validated for bulk structure prediction by comparing the RDFs obtained from NNIP with those calculated from AIMD simulation (training data). Here, the peak positions in partial RDFs obtained from simulations were also compared with the corresponding experimental peak positions reported by Ganster et al. [7]. **Figure 8(a)** shows the total RDF comparison between NNIP and AIMD as obtained using OVITO [63] for 10 units PAN system simulated at the experimental density of ~1.17 g/cc. As observed, the total RDF from both simulation methods agree well, ensuring accurate prediction of bulk PAN structure using NNIP. The magenta dashed lines mark the experimental peak positions for the corresponding atomic pairs for the atactic PAN synthesized using a radical polymerization [7]. The simulated peak positions for AIMD and NNIP-based MD simulations are in excellent agreement with the experimental values marked in **Figure 8(a)**.

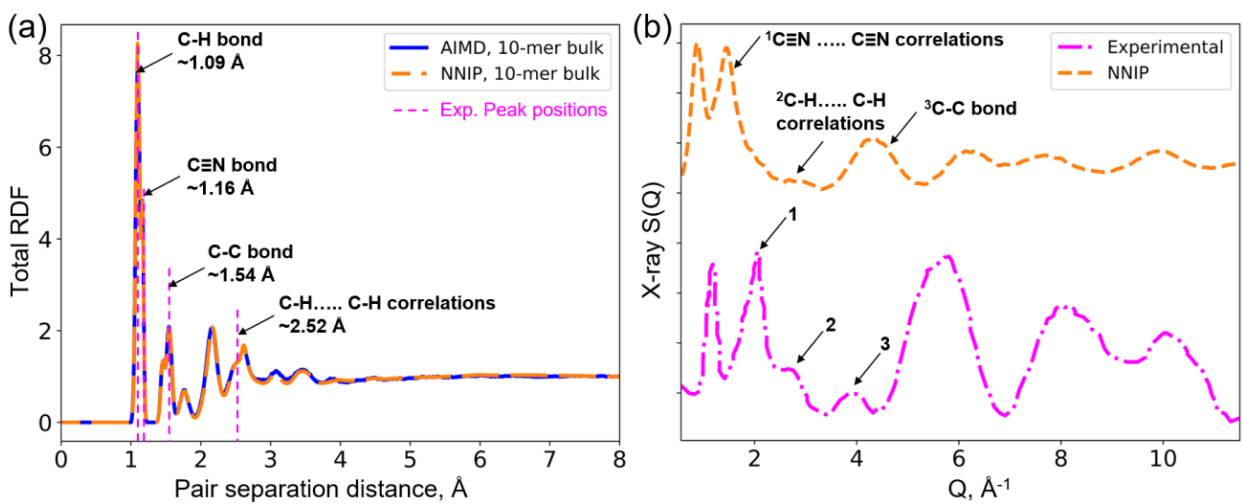

**Figure 8.** (a) Radial distribution analysis to validate NNIP against AIMD simulations for 10-mer bulk and (b) Comparison of x-ray structure factor (S(Q)) for 320-mer bulk system studied using NNIP. Experimental x-ray data is obtained from Ganster et al. [7].

Further, the simulated microstructures were analyzed to evaluate x-ray structure factor (**Equation S4**) for bulk PAN, which was compared with the experimental x-ray S(Q) (**Figure**

**8(b))**. Here, the correlations in x-ray S(Q) [64] at ~1.85 Å$^{-1}$ is due to the pendant group, i.e., C≡N….. C≡N interactions. In a bulk system, a small peak at ~2.85 Å$^{-1}$ is observed due to C-H…..C-H correlations, i.e., C-C next-nearest-neighbor. At Q ~ 4.35 Å$^{-1}$, a peak representing C-C bond in PAN chains is observed in the simulated x-ray S(Q). A sharp peak at nearly ~0.85 Å$^{-1}$ in simulated structure factor corresponds to intermediate range structures formed at more-than 7-8 Å radial distances in real space. Similar observations were made by Ganster et al. in their experimental x-ray S(Q) measured for atactic PAN [7] (**Figure 8(b)**). Here, the observed deviation in S(Q) peak positions and intensities could be due to the observed density differences between simulated and experimental S(Q).

A comparison of the effect of simulated X-ray S(Q) obtained from systems comprising different chain molecular weights is shown in **Figure 9**. Here, similar features were observed in all the bulk systems, irrespective of the molecular weight of the PAN chains. The corresponding real space correlation functions (partial RDF) are shown and compared in **Figure S4.**

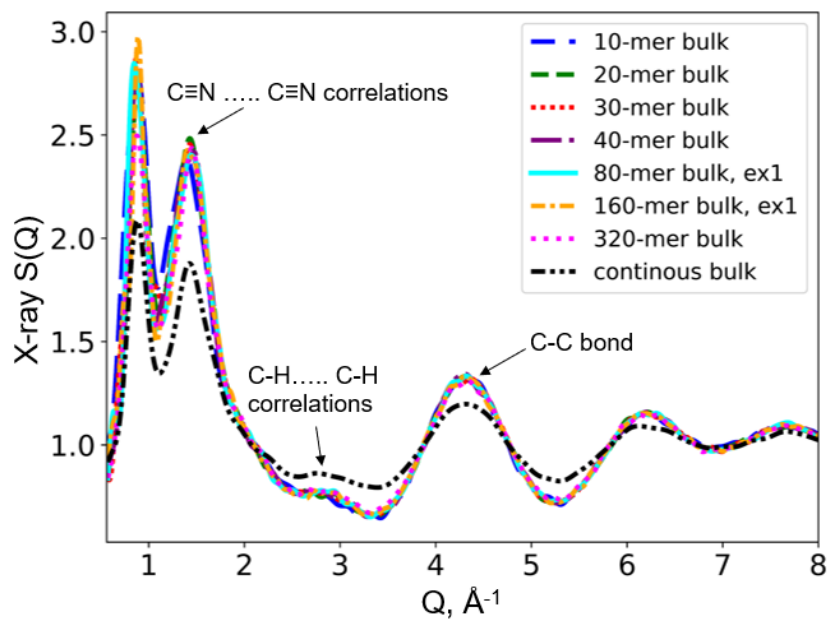

**Figure 9.** Comparison of X-ray structure factor for n-mer bulk systems studied using NNIP.

Upon validating the PAN structure using AIMD and experimental RDFs and S(Q) data, the NNIP was further employed to study C≡N….H hydrogen bonding, relaxation behavior, and elastic properties of bulk PAN from different n-mer chains.

*H-bonding*

Bulk PAN simulated microstructures were further analyzed for H-bonding interaction approximated using C≡N….H coordination. Here as well, the C≡N….H coordination was obtained using **Equation 1**, where $r_{cut}$= 2.85 Å was chosen based on the first minimum in H-N partial $g(r)$ calculated for bulk PAN (**Figure 10(a)**). To compare the H-bonding interaction among different n-mer systems, the calculated value of C≡N….H coordination for each system was normalized by the number of n-mer units. **Figure 10(b)** compares the H-bonding per n-mer as calculated for bulk systems. Here, in contrary to the trend of increase in the H-bonding/n-mer in the chain as shown in **Figure 6(b)**, H-bonding/n-mer the in bulk is nearly constant across the n-mer count. It is because even though intrachain interactions are limited due to prohibitively small chain size, the interchain C≡N….H coordination can contribute to the higher values of net H-bonding. **Figure 11(b)** shows a snapshot of the H-bond number of 3 mediated by the interchain C≡N…..H coordination.

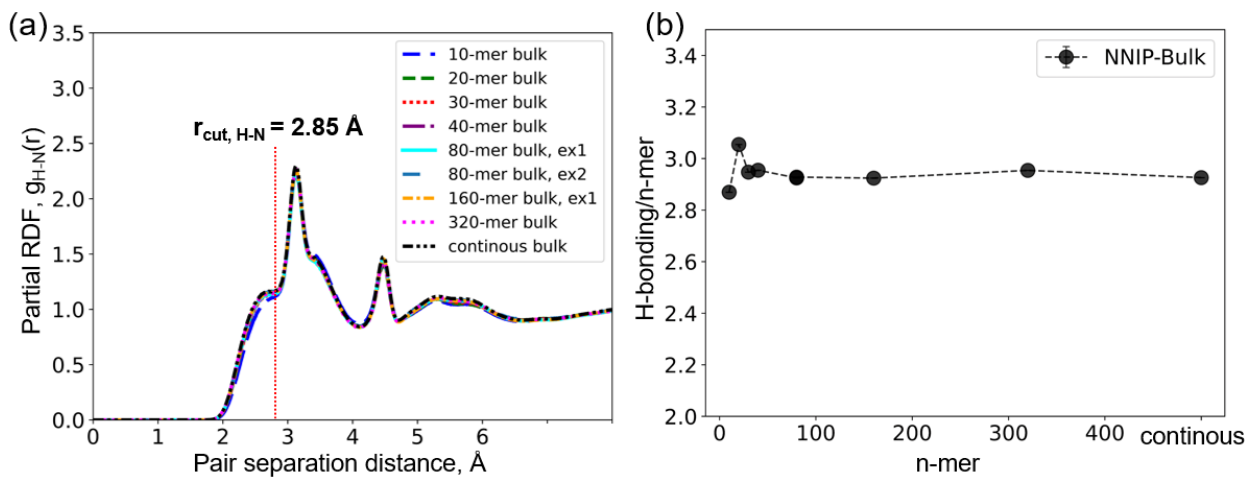

**Figure 10.** (a) C≡N…..H partial RDF evaluated for bulk and (b) H-bonding (C≡N…..H coordination) predicted by NNIP for bulk PAN.

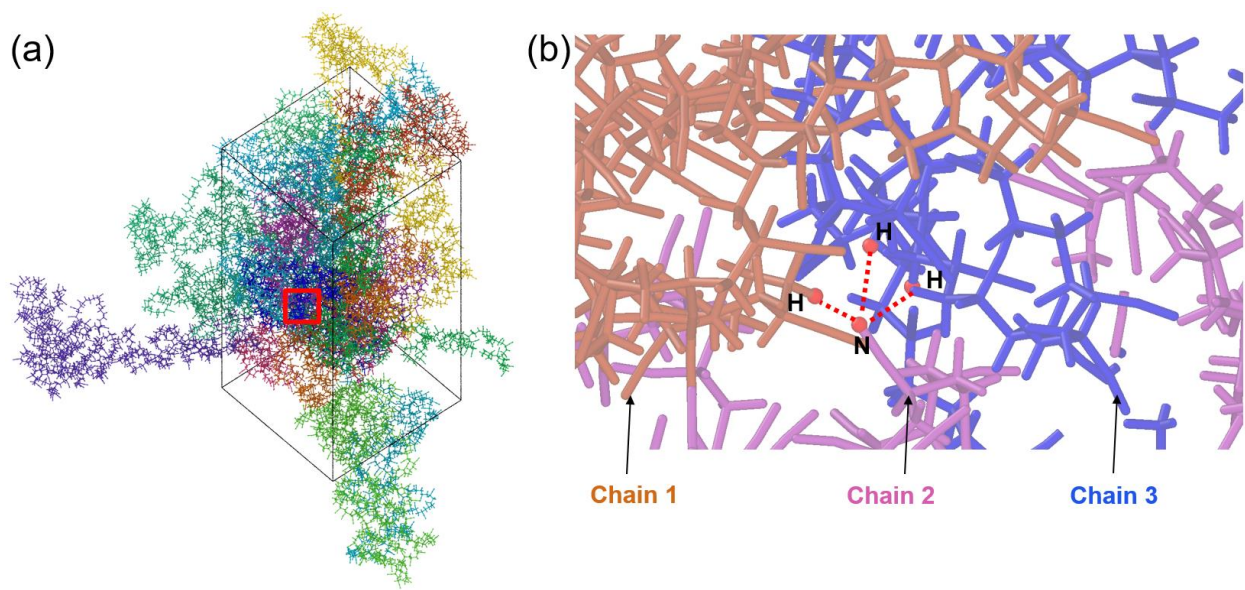

**Figure 11.** (a) Packing of PAN chains in the 320-mer system and (b) snapshot of C≡N…..H coordination, highlighting 3 H-bonds. Here, C≡N from Chain 2 forms coordination with H atoms from Chain 1 and Chain 3. Different chains are color-coded using OVITO [63].

*Dipole Relaxation Behavior of Bulk PAN*

The dipole ACF function for 10-mer, 160-mer, and continuous chain bulk systems was evaluated and compared from 14 ns, 10 ns, and 13 ns NNMD simulations, respectively (shown in **Figure 12**). Although a thorough quantitative understanding of the relaxation time could be achieved given even longer simulations (milliseconds) are performed, nevertheless, a trend in pendant group relaxation behavior can be attained. As noticed in **Figure 12**, a comparatively faster dipole decay is observed for the 10-mer system, while both the 160-mer and continuous chain bulk systems show rather slower dipole relaxation behavior. Herein, the observed slow dipole relaxation response for a larger n-mer system could be attributed to additional degrees of freedom and more chemical transformation events in the system as the chain molecular weight increases. As in PAN, contrary to the small polar molecules, the steric constraints introduced by the backbone chain and other substituents may cause a high correlation in the orientation of large numbers of dipolar side groups, which may then be able to move only in some cooperative fashion. This implies that any dipole may be capable of relaxing by way of more than one mode of motion, necessitating longer simulation times and multiexponential fitting to precisely quantify the overall relaxation behavior.

The experimental values obtained from dielectric measurements for dipole relaxation time for atactic PAN reported for temperature range 30°C-150°C are between 5 millisecond (ms)-50 ns [36]. It should be noticed that in the referenced experimental study, as the samples prepared for dielectric measurements comprised crystalline phase in addition to the amorphous regions, the experimental relaxation time is expected to be higher than that obtained from amorphous PAN microstructure simulations [36]. Yet, in order to obtain the comprehensive and accurate relaxation behavior for estimating relaxation time at 300 K, tens of ms long simulations need to be performed to reach a vanishingly small ACF (< 1%), which is still prohibited by the computational cost for bulk systems. As such, this study qualitatively demonstrated the effect of chain molecular weight on the C≡N dipole relaxation behavior, suggesting that the relaxation time increases upon increasing the chain molecular weight and may eventually approach a constant value upon further increase in the chain size.

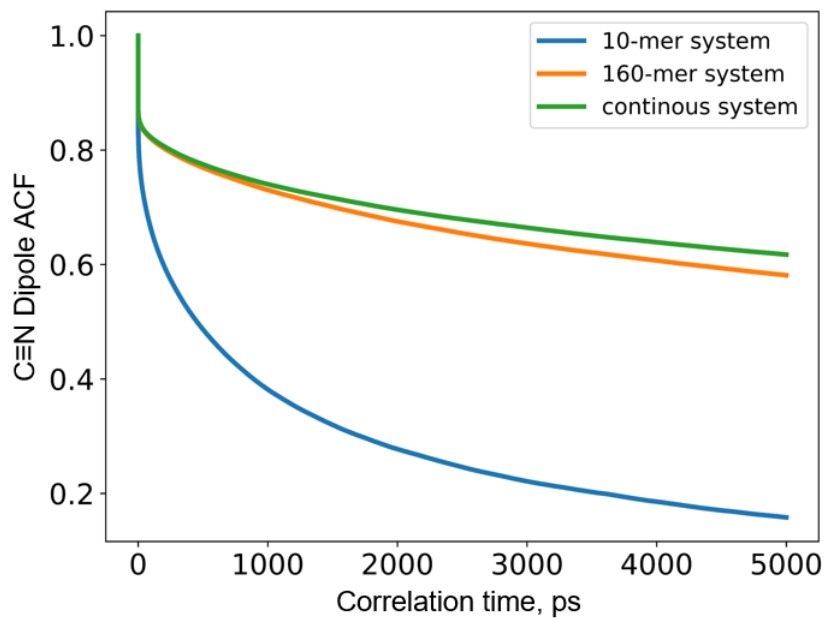

**Figure 12.** C≡N dipole relaxation behavior predicted by NNIP for 10-mer, 160-mer, and continuous chain PAN bulk systems.

*Elastic properties: Experiments and NNMD*

Experimental results from scattering and tensile testing are shown in **Figure 13**, which reveals the primary scattering feature defined by the intersegmental packing within the crystal lattices. Fibers consists of strained polymer chains aligned along the fiber axis. These strained chains form packed crystalline domains within the orientated amorphous polymer strands. Misorientation of polymer strands caused by thermal relaxation induces distributed alignment of polymer chains and their packed crystal domains. The chain alignment direction, parallel to the fiber axis, was defined as the reference axis and resulted in Herman's orientation factor values of 0.382, 0.402, and 0.417 (calculated using **Equation 4, 5**) for the steam relaxed, as-spun, and steam stretched samples, respectively. While samples did show an increase in HOF in correlation with processes which would impart more or less alignment (i.e. alignments of stream relaxed < as spun < steam stretched), all values still showed a considerable degree of alignment. The relatively low HOF for steam relaxed samples is indicative of the strong secondary interactions brought about by the nitrile group limiting thermal relaxation.

After characterization, the fiber samples were tested using a Favimat single-filament tensile testing machine. Results are shown below. Results depict the stress-strain profile of the as-spun (black), steam relaxed (red) and steam stretched (blue). From the stress-strain profiles, modulus was calculated using linear fit between 0.2 and 0.6% strain. As can be seen in the figure, the as-spun samples generally fell in between steam-stretched and steam-relaxed samples, indicating that the post-processing did influence the final expressed properties. The greater spread of data from the steam stretched and steam relaxed samples are attributed to uneven loading on the filaments within the tow during the process. Despite the increased spread of the post-processed fibers, the average modulus was 6.96±0.76, 7.86±0.34, and 8.55±0.62 GPa for the steam relaxed, as spun, and steam stretched samples, respectively. The increasing modulus with increased alignment is commonly reported for aligned materials [65].

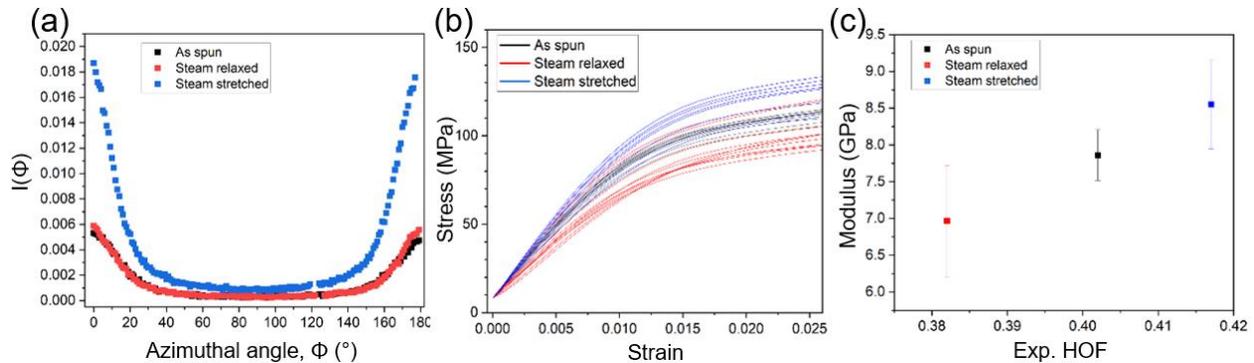

**Figure 13.** Structure-property measurements of as-spun, steam-relaxed, and steam-stretched materials. Steam-stretched samples displayed the highest alignment and modulus. The average modulus was 6.96±0.76, 7.86±0.34, and 8.55±0.62 MPa for steam-relaxed, as-spun, and steam-stretched samples.

Further, to determine the elastic properties from NNMD simulations, we considered the bulk systems comprising polymer chains with 10-320 monomer units as well as the continuous chain system. The elastic properties were evaluated through tensile testing approach, i.e., by stretching the NNMD-equilibrated samples along the x-direction at a strain rate of $1 \times 10^9$ s$^{-1}$. Before loading, all the samples were at corresponding equilibrium densities (values shown in **Figure 7**) to achieve near-zero residual stresses in the system. During x-direction loading, the simulation box was allowed to relax in lateral (y and z) directions to allow for Poisson's contraction using Nosé-Hoover barostat/thermostat [49] maintaining a lateral pressure of 1 atm and a temperature of 300 K. The stress-strain relationship was evaluated using the longitudinal stress component evaluated within 1% strain (fit shown in **Figure S5**). **Figure 14** shows the corresponding mechanical properties evaluated from the loading simulations. The E values of 2.15-3.37 GPa were evaluated for PAN systems comprising chain sizes from 10-mer to 320-mer. With exception to the continuous chain system, the elastic moduli values are observed to be nearly insensitive (within uncertainty) to the number of repeat units. The Young's modulus for continuous chain system was calculated to be E=5.24±0.68 GPa.

To gain more insight into the observed deviation and trend in modulus values obtained in simulations, the Hermans orientation factor [56] is established as a parameter (**Equation 3**). It has been previously employed to correlate the stiffness of polymers and their structural alignment [41], [42], [65]. Here, the bulk PAN microstructures in NNMD simulations were characterized to identify C-C backbone alignment in each system that can be further used to obtain HOF value for the backbone chain with respect to the loading axis (here, x-axis). **Figure 15** shows the backbone carbon chain and $\theta_x$ angle distributions for 10-mer, 160-mer, and continuous chain bulk systems, which were used for calculating HOF values using **Equation 3**. The correlation in elastic modulus and HOF values for all simulated (accounting for backbone alignment) and measured systems (accounting for crystallite alignment) are shown in **Figure 14(b)**. It can be observed that the simulated values of elastic moduli are correlated with the degree of backbone chain alignment. As such, for systems with comparable elastic moduli values (10-mer to 320-mer systems), the C-C backbone orientation is also comparable as represented by their HOF values=-0.0455-0.0323 (near random backbone orientation). Further, as the backbone alignment is better for a continuous chain system (HOF~0.127), the elastic modulus of PAN in this case was higher (E=5.24±0.68 GPa). This suggests that the elastic properties of the PAN bulk system are less sensitive of the chain molecular weight, while having a large influence of C-C backbone orientation on the elastic modulus. As pointed out earlier in this section, a similar correlation between E values and measured HOF was found in the experimental measurements.

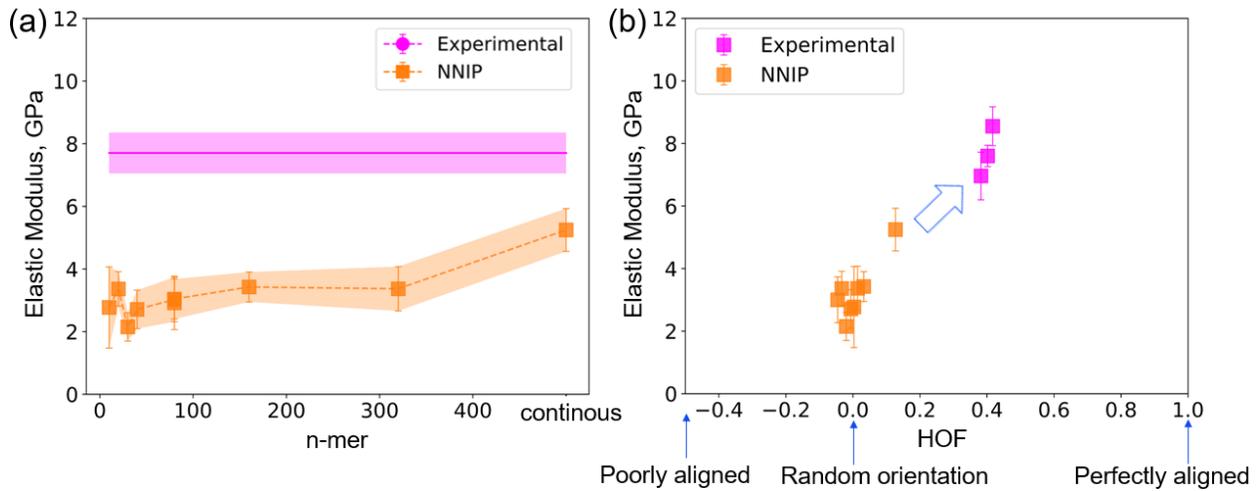

**Figure 14.** (a) NNIP predictions of PAN elastic modulus and (b) correlation between elastic modulus and simulated (**Equation 3**) and measured (**Equation 4 and 5**) HOF values. The categorization marked on x-axis is based on HOF value calculated with respect to the loading axis (here, x-direction).

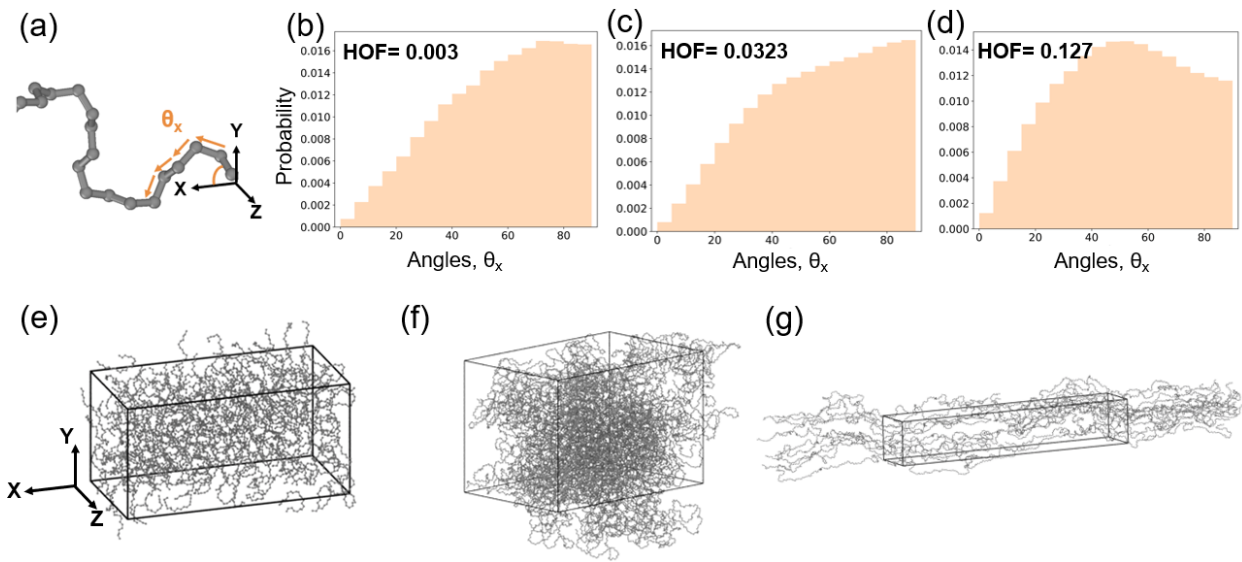

**Figure 15.** (a) C-C backbone segment orientation angles with respect to loading x-axis to obtain Hermans orientation factor (HOF), (b-d) distribution of $\theta_x$ for 10-mer, 160-mer, continuous chain system, and (e-g) show C-C backbone in these systems.

**DISCUSSION AND CONCLUSION**

In this work, we demonstrated the applicability of machine learning for studying the structure, dynamics, and macroscale properties of PAN polymers. We developed an NNIP for PAN simulation by leveraging AIMD simulations of small PAN systems comprising 10- and 20-monomer units, which in turn allowed us to successfully simulate much larger PAN chains (n-mers > 300) and their bulk configurations. Using the developed NNIP validated with new and available experimental data, we unraveled the structural complexities of PAN at the molecular level, namely from an H-bonded folded state to an unfolded, maximally-aligned state, and elucidated their dynamical behavior (the C≡N dipole rotational dynamics) and the mechanical properties as a function of the PAN chain length.

For isolated/individual PAN chains, the effect of n-mer units (i.e., chain length) on the radius of gyration, Flory's random coil hypothesis, and H-bonding were studied using both NNIP and ReaxFF parameters. It was found that as n-mer count increases, the chain radius of gyration increases and eventually reaches a plateau, as the enhanced C≡N dipole-dipole interactions upon the increase in the PAN chain size led to more conformational folding within the chain. A similar trend was observed for H-bonding between the C≡N groups and H atoms in the chains. For the bulk PAN systems, as interchain C≡N dipole interactions occur in addition to intrachain C≡N dipole interactions; therefore, the C≡N…. H coordination stays nearly independent of the chain size. Nevertheless, the effect of the chain length was observed on the C≡N dipole relaxation behavior, where the relaxation time was found to increase upon increasing the chain length. The relaxation behavior eventually saturates upon further increase in the chain length. Similarly, the density of PAN reaches a plateau after 40-mer. The NNIP predicted values for density ($\rho_{NNIP}$ ~1.1±0.1 g/cc) are within 7% of the reported experimental values in the literature ($\rho_{exp}$~1.19±0.025 g/cc). Furthermore, the mechanical properties of the PAN bulk system (E=2.15-5.24 GPa) were found to be independent of chain molecular weight, while a significant influence of C-C backbone orientation on the elastic modulus was observed by comparing Hermans orientation factor (HOF) for each system. It was found for the systems with ~0 HOF value (near random backbone orientation), E values vary between 2.15-3.37 GPa. Meanwhile, as the backbone alignment improves (HOF~0.127), the elastic modulus of PAN increases (E=5.24±0.68 GPa). A similar observation was made in experimental measurements, where the E values increase from 6.96 GPa to 8.55 GPa as HOF increases from 0.382 to 0.417. As such, both experiments and NNIP simulations established that the elastic moduli values are directly correlated with the degree of the PAN chain alignment.

Lastly, the developed NNIP was shown to be more efficient than the conventionally used ReaxFF, while sustaining the *ab initio* accuracy across scales. Nevertheless, it should be acknowledged that ReaxFF has excellent capabilities to accommodate molecular reactions. Here, given the NNIP potential's efficiency, employing NNIP will be desirable to treat bond-breaking or bond-forming phenomena in macromolecules and materials [66], [67], [68]. As such, the established workflow for NNIP training in this study can be straightforwardly extended as done in Ref. [66], [67], [68] for future studies of a broad range of reactive polymeric systems, thus, paving the way for understanding and establishing their molecular structure-property relationships.

## ASSOCIATED CONTENT

**Supporting Information**.

The following files are available free of charge.

Supplementary information file: Detailed information on NNIP training, simulated x-ray S(Q) calculation, PAN configuration snapshots, Radial distribution plots, Stress-strain relationship curves fitted to calculate moduli.

The developed NNIP potential energy surface will be made available at
https://github.com/rajnichahal/.

## AUTHOR INFORMATION

**Corresponding Authors**

**Author Contributions**

RC led and performed all MD simulations, NNIP development and deployment, including subsequent computational analyses. MT, LTK, JTD, and AKN conducted experimental design and measurements. AS conducted DeepMD installation on the Summit supercomputer at ORNL. SR designed and administered the research project and partially assisted with the analysis. The original manuscript draft was prepared by RC, SR, MT, and LTK. All authors contributed to proofreading the manuscript.

There are no conflicts to declare.

**ACKNOWLEDGEMENT**

This research at ORNL was sponsored by the US Department of Energy (DOE)'s Office of Energy Efficiency and Renewable Energy, Hydrogen and Fuel Cell Technologies Office Program (Award # DE-LC-000L83). ORNL is managed by UT Battelle, LLC, for DOE under contract DE-AC05-00OR22725. Experimental wide-angle x-ray scattering work (L.T.K and M.D.T) was supported by the US Department of Energy, Office of Science, Basic Energy Sciences, Materials Sciences and Engineering Division [FWP# ERKCK60].  This research used resources of the Oak Ridge Leadership Computing Facility at the Oak Ridge National Laboratory, which is supported by the Office of Science of the U.S. Department of Energy under Contract No. DE-AC05-00OR22725. A part of this research used resources of the National Energy Research Scientific Computing Center (NERSC), a DOE Office of Science User Facility supported by the Office of Science of the U.S. Department of Energy under Contract No. DE-AC02-05CH11231 using NERSC award BES-ERCAP0027122. This research also used resources of the Oak Ridge Leadership Computing

Facility at the Oak Ridge National Laboratory, which is supported by the Office of Science of the U.S. Department of Energy under Contract No. DE-AC05-00OR22725. We would like to thank Vyacheslav Bryantsev and Alex Ivanov at ORNL for sharing these resources to perform this work.

**For Table of Contents Only**

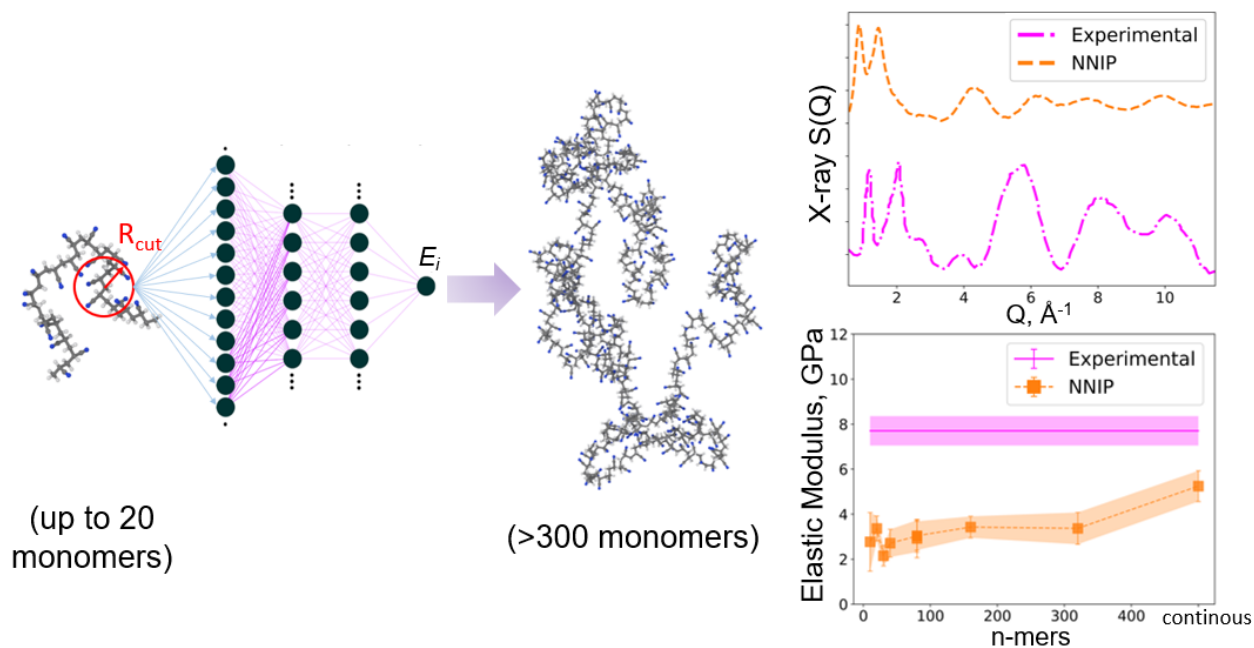